\documentclass[letterpaper,10pt]{article} 
\usepackage{osameet3} %% use version 3 for proper copyright statement

\usepackage{amsmath,amssymb}
\usepackage{float}
\usepackage[colorlinks=true,bookmarks=false,citecolor=blue,urlcolor=blue]{hyperref} %pdflatex
\usepackage{blindtext, graphicx, tabularx, multicol, lipsum, eqnarray,amsmath,chemformula,multirow,threeparttable, subfigure}
\usepackage{algorithm}
\usepackage[noend]{algpseudocode}
\usepackage{caption}

\begin{document}

\title{CHAMP: Coherent Hardware-Aware Magnitude Pruning of Integrated Photonic Neural Networks}

\author{Sanmitra Banerjee\textsuperscript{1}, Mahdi Nikdast\textsuperscript{2}, Sudeep Pasricha\textsuperscript{2}, Krishnendu Chakrabarty\textsuperscript{1}}
\address{\textsuperscript{1}ECE Dep., Duke University, Durham, NC, USA \textsuperscript{2}ECE Dep., Colorado State University, Fort Collins, CO, USA}
\email{sanmitra.banerjee@duke.edu, mahdi.nikdast@colostate.edu, sudeep@colostate.edu, krish@duke.edu}

%% Uncomment the following line to override copyright year from the default current year.
\copyrightyear{2022}

\begin{abstract}
We propose a novel hardware-aware magnitude pruning technique for coherent photonic neural networks. The proposed technique can prune 99.45\% of network parameters and reduce the static power consumption by 98.23\% with a negligible accuracy loss.
\end{abstract}
\ocis{(200.4260) Neural networks; (250.5300) Photonics integrated circuits}
%\vspace{-0.2in}
\section{Introduction}
\begin{figure}[H]
  \centering
  \subfigure[A 4$\times$4 linear multiplier in an SC-IPNN]{
\includegraphics[width=.375\textwidth]{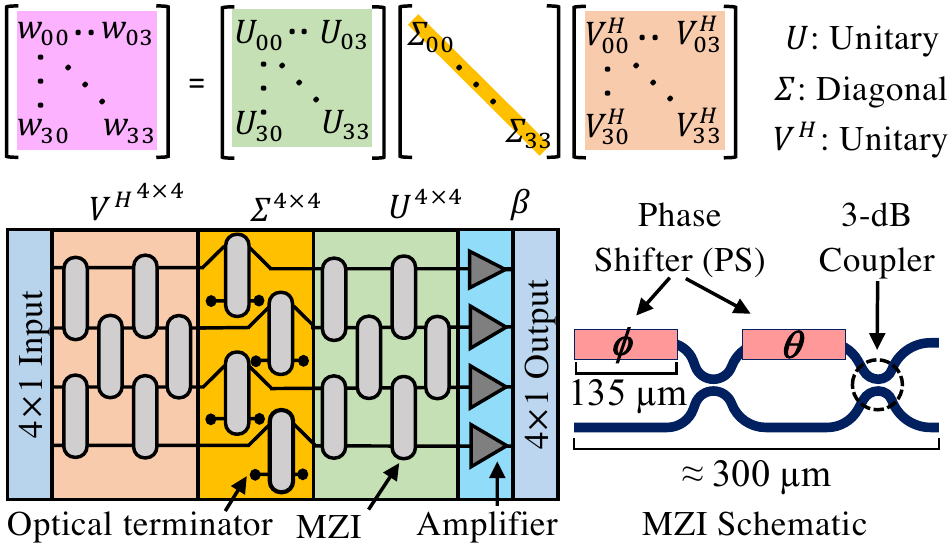}
}\hspace{-0.02in}%
\subfigure[Bidirectional many-to-one mapping]{
\includegraphics[width=.315\textwidth]{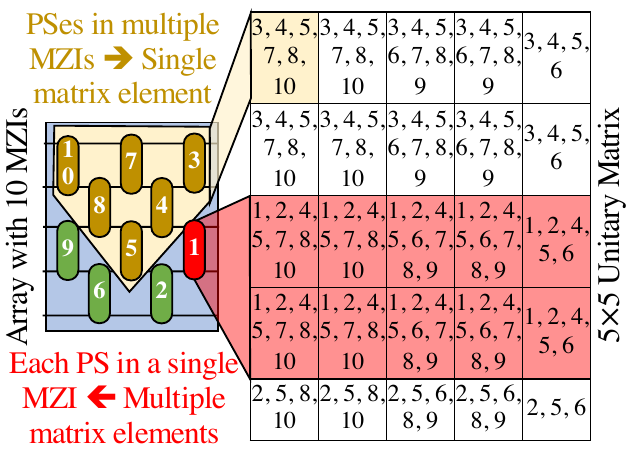}
}\hspace{-0.04in}%
\subfigure[Matrix sparsity versus PS sparsity]{
\includegraphics[width=.285\textwidth]{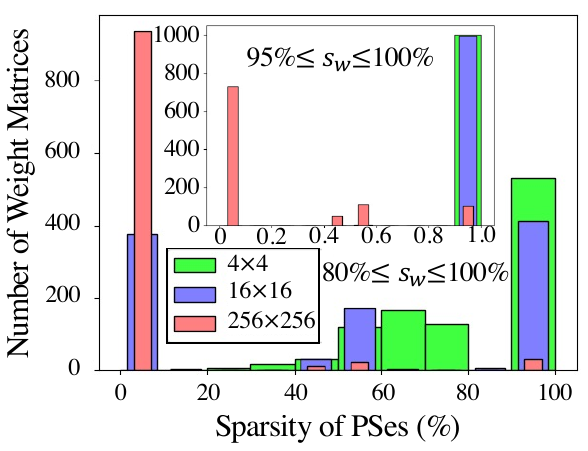}
}
\vspace{-1em}
  \caption{(a) A linear multiplier implemented based on SVD and using MZI arrays (MZI dimensions are obtained from \cite{shokraneh2020theoretical}). (b) An illustration of the bidirectional many-to-one mapping between the elements of a 5$\times$5 unitary matrix and the mapped MZI array. The numbers in each cell of the unitary matrix denote the MZIs that affect the corresponding matrix element. (c) Histogram of the sparsity of PSes (percentage of zero phase angles) in the mapped MZI arrays for 3000 randomly generated weight matrices of different dimensions (1000 of each dimension) with sparsity $s_w$, where 80\%$\leq s_w \leq$100\%. The inset shows a similar plot for 95\%$\leq s_w \leq$100\%. }
  \vspace{-1.5em}
  \label{fig1}
\end{figure}

\vspace{-0.05in}
Coherent integrated photonic neural networks (C-IPNNs) promise ultra-fast and ultra-low-energy linear multipliers for emerging artificial intelligence (AI) accelerators. C-IPNNs based on singular value decomposition (SVD), referred to as SC-IPNNs in this paper, factorize a linear multiplier into one diagonal and two unitary matrices, each of which can be implemented using an array of Mach--Zehnder interferometers (MZIs). During the training, the phase angles on each MZI ($\phi$ and $\theta$ in Fig. \ref{fig1}(a)) are adjusted using stochastic gradient descent to minimize the overall training loss. Nevertheless, SC-IPNNs suffer from a large footprint and high static power consumption. In particular, SC-IPNNs employ phase shifters (PSes)---implemented often using thermo-optic phase-shift mechanisms---where the phase change ($\Delta\phi$) is directly proportional to the tuning power consumption ($P$) and the PS length ($L$): $\Delta\phi\propto P\cdot L$~\cite{jacques2019optimization}. To maintain the phase shifts, the tuning power is consumed throughout inferencing and can range up to 25 mW/$\pi$ \cite{harris2014efficient}. In addition, the underlying PSes in the MZI devices in SC-IPNNs (see $\phi$ and $\theta$ in Fig. \ref{fig1}(a)) account for a significant portion of network footprint (e.g., up to $\approx$90\% of a single MZI footprint designed in \cite{shokraneh2020theoretical}). Moreover, it was shown that uncertainties in PSes, especially in those with high phase angles, can cause up to 70\% loss in SC-IPNN accuracy~\cite{banerjee2021optimizing}. A potential solution to the aforementioned problems is to prune the redundant PSes and minimize the phase angles in the network. Prior attempts at pruning SC-IPNNs use a software-only approach where a trained network is first pruned and the resultant sparse weight matrices are subsequently mapped to the MZI arrays using SVD. However, due to the complex mapping between the weights and the MZI arrays in SC-IPNNs (see Fig. \ref{fig1}(b)), sparse weight matrices often lead to non-sparse MZI phase settings (and vice versa). Consequently, software-only pruning is inefficient in SC-IPNNs and imposes significant accuracy losses. To enable efficient pruning in SC-IPNNs, we propose the first hardware-aware pruning technique for SC-IPNNs, called CHAMP. As we will show, in a representative SC-IPNN with 155,268 PSes, CHAMP can prune up to 74.86\% of PSes with no accuracy loss and up to 99.45\% of PSes with an accuracy loss of less than 5\%. These correspond to a 46.05\% and 98.23\% reduction in static power consumption, respectively. Additionally, if the redundant PSes are removed (rather than being power-gated), the resultant SC-IPNN demonstrates significantly smaller footprint---which in turn will help reduce the dynamic power consumption (the analysis of which is beyond the scope of this paper)---and higher immunity to uncertainties in phase angles. 
\vspace{-0.2in}

\section{CHAMP: Proposed Hardware-Aware Magnitude Pruning Framework}\vspace{-0.05in}
In hardware-unaware pruning techniques, a fraction of the weights in each neural network layer---typically those with a smaller magnitude---are clamped to zero. Then, the network is retrained (i.e., fine-tuned) to recover the accuracy while ensuring that only the non-zero weights are updated. However, the mapping of the sparse weight matrices obtained using hardware-unaware pruning approaches to MZI arrays may not necessarily lead to sparse PSes. Fig. \ref{fig1}(c) shows that several randomly generated sparse weight matrices are \textit{not} mapped to sparse PSes (especially true for larger weight matrices). The discrepancy between the sparsity of the weight matrices and their corresponding PS mappings is due to the \textit{bidirectional many-to-one association (BMA)} between the elements of the weight matrix and the phase angles. Each element of the weight matrix of a linear layer in SC-IPNNs is mapped to multiple phase angles and each phase angle in an MZI array affects multiple matrix element, as shown in Fig. \ref{fig1}(b). Prior work using hardware-unaware pruning showed that no more than 30\% (45\% for non-SVD-based C-IPNNs) of the phase angles can be pruned without a significant ($\approx$10\%) accuracy loss \cite{gu2020towards}. \par
\begin{figure}[t]
  \centering
  \subfigure[Interlinking between one-shot and iterative pruning approaches in CHAMP]{
\includegraphics[width=.61\textwidth]{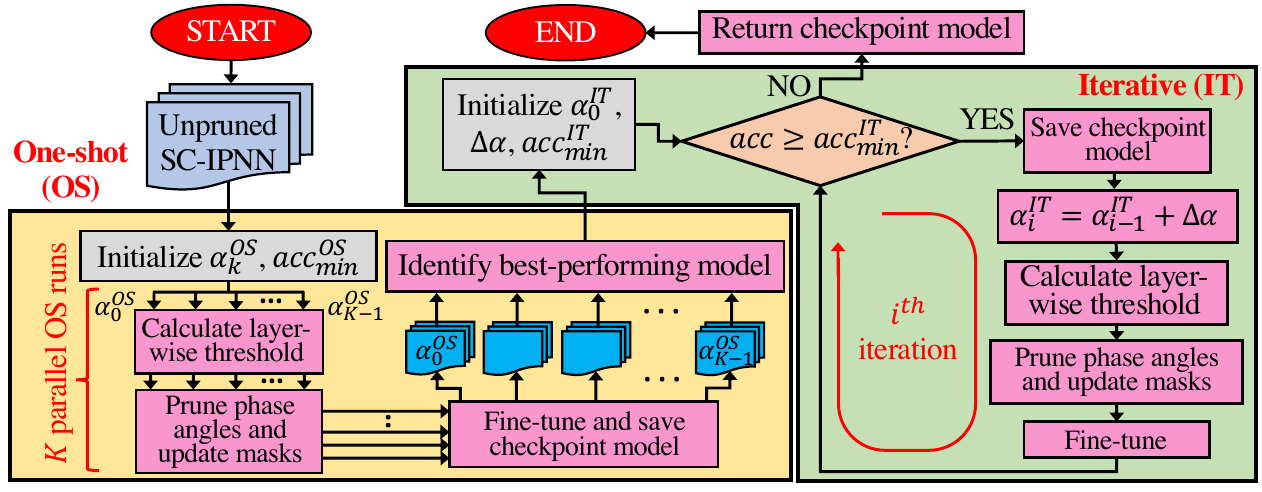}
}\hspace{-0.01in}%
\subfigure[One-shot (OS) magnitude pruning]{
\includegraphics[width=.36\textwidth]{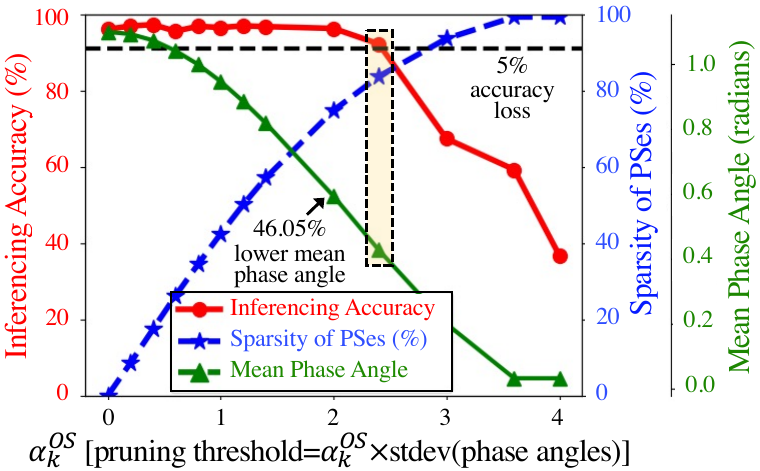}
}
\vspace{-1em}
  \caption{(a) Block diagram highlighting the hybrid (one-shot and iterative) pruning approach in CHAMP. (b) Fine-tuned inferencing accuracy, PS sparsity, and mean phase angle for one-shot magnitude pruning for different $\alpha_{k}^{OS}$. The yellow rectangle shows the best-performing (high sparsity and low accuracy loss) one-shot-pruned model.}
  \vspace{-2em}
  \label{fig2}
\end{figure}

The key difference between the existing pruning methods and CHAMP lies in the training approach. Instead of mapping the software-trained weight matrices to phase angles, we use a \textit{photonic hardware-aware} approach where, during backpropagation, the phase angles (and not the weight matrix elements) are adjusted based on the computed gradients. Photonic hardware-aware software training offers more control on the phase angles during training and is essential for efficient SC-IPNN pruning. Fundamentally, in magnitude pruning, the weights (phase angles here) with magnitude smaller than a threshold are set to zero. Next, the pruned network is retrained to recover the lost accuracy while clamping the pruned phase angles to zero. A common approach to determine the threshold for the phase angles in an MZI array is to consider a factor---say $\alpha$---of the standard deviation of the non-zero phase angles in the array. The higher the value of $\alpha$, the more aggressive is the pruning. We also maintain a binary mask for each phase angle in the MZI layer; an element of the mask is 0 (1) if and only if the corresponding phase angle is zero (non-zero). During backpropagation, the computed gradient of each phase angle is multiplied with its corresponding mask element, thus ensuring that the zero phase angles are not perturbed. Magnitude pruning can be performed in a one-shot or iterative manner. In the one-shot approach, all phase angles below a threshold are pruned in a single step after which retraining (a.k.a. fine-tuning) is performed. In the iterative approach, $\alpha$ (and, in turn, the pruning rate) is increased over multiple steps with each step followed by fine-tuning. \par

Fig. \ref{fig2}(a) shows a block diagram of the CHAMP framework. We use a hybrid approach where the faster one-shot (OS) pruning is used to quickly ramp up the sparsity of the PSes, and then the iterative (IT) approach is employed to increase the sparsity further while maintaining a high-enough model accuracy. The inputs to the OS approach include the trained SC-IPNN, the minimum acceptable inferencing accuracy ($acc_{min}^{OS}$), and a set of $K$ $\alpha$'s to determine the thresholds ($\alpha_{k}^{OS}$, $k=$~0, 1, $\dots$, $K-1$). The $K$ OS runs are mutually independent and can be executed in parallel. Out of the $K$ OS-pruned models, the best-performing one (with maximum sparsity and accuracy greater than $acc_{min}^{OS}$) is considered for IT pruning. The inputs in the IT approach include the initial $\alpha$ ($\alpha_{0}^{IT}$), the step-increment in $\alpha$ ($\Delta \alpha$), and the minimum acceptable accuracy ($acc_{min}^{IT}$). The $\alpha$ for the $i^{th}$ iteration ($i\geq$~1) is given by $\alpha_{i}^{IT}=\alpha_{i-1}^{IT}+\Delta \alpha$. The IT approach terminates when the inferencing accuracy becomes less than $acc_{min}^{IT}$; in this case, the checkpoint model saved in the previous iteration is returned as the sparse SC-IPNN. Also, in different pruning runs, we consider the same $\alpha$ ($\alpha_{k}^{OS}$ and $\alpha_{i}^{IT}$) for each SC-IPNN layer. However, the threshold phase angle (given by $\alpha\times$std. dev. of non-zero phase angles in a layer) can differ from layer to layer.\vspace{-0.05in}

\section{Results and Discussion}
\begin{figure}[t]
  \centering
  \subfigure[Iterative (IT) magnitude pruning]{
\includegraphics[width=.385\textwidth]{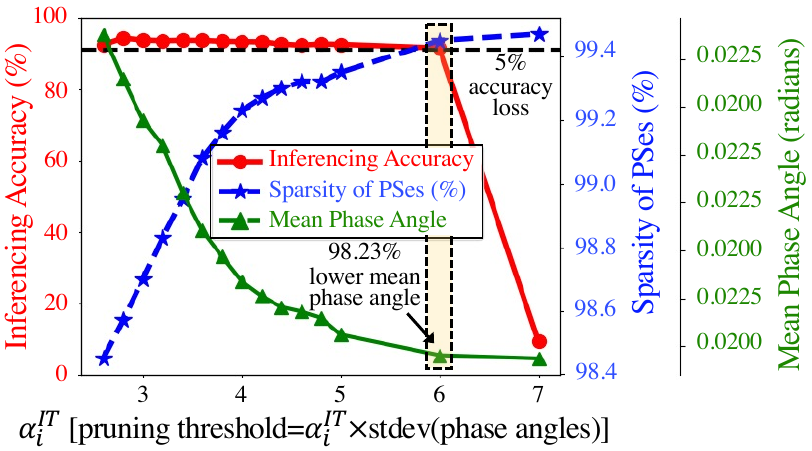}
}\hspace{-0.01in}%
\subfigure[Histogram of phase angles]{
\includegraphics[width=.289\textwidth]{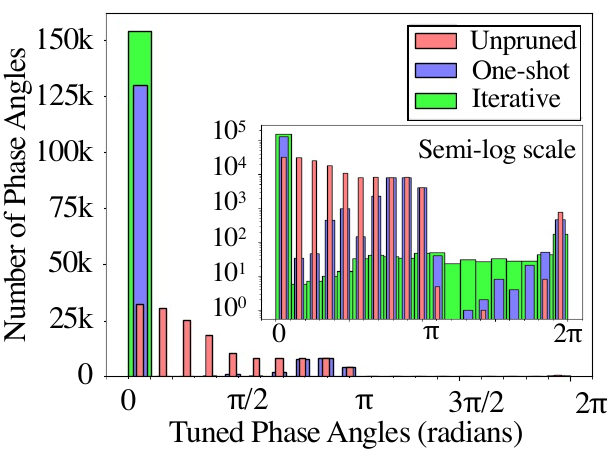}
}\hspace{-0.05in}%
\subfigure[Accuracy under phase uncertainties]{
\includegraphics[width=.295\textwidth]{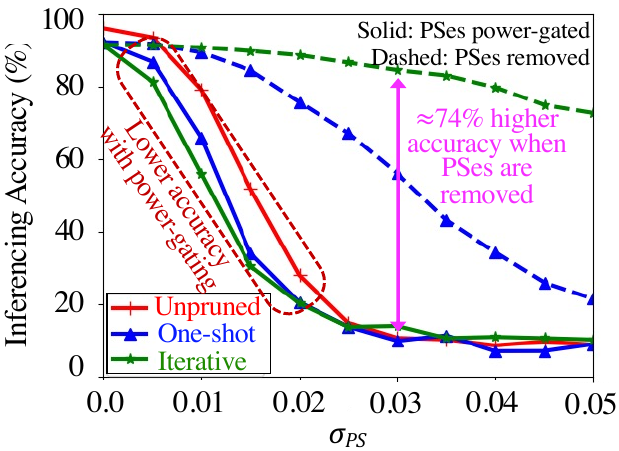}
}
\vspace{-1em}
  \caption{(a) Fine-tuned inferencing accuracy, PS sparsity, and mean phase angle for iterative pruning for different $\alpha_{i}^{IT}$. Yellow rectangle: best-performing IT-pruned model. (b) Histogram of the tuned phase angles in unpruned, best-performing one-shot-pruned, and best-performing iterative-pruned models. (c) Accuracy of the unpruned, best-performing one-shot-pruned, and best-performing iterative-pruned models under random phase uncertainties.}
  \vspace{-2.5em}
  \label{fig3}
\end{figure}

\vspace{-0.05in}
To demonstrate the performance of CHAMP, we consider a case study of a fully connected feedforward SC-IPNN with two hidden layers (with 256 and 100 neurons) and 155,268 PSes, implemented using the Clements design \cite{clements2016optimal}. We train the network on the MNIST dataset; each real-valued image is converted to a complex feature vector of length 64 using a method based on fast Fourier transform \cite{banerjee2020modeling}. The nominal inferencing accuracy of the unpruned network is 96.16\%. Fig. \ref{fig2}(b) shows the simulation results using the OS approach for different values of $\alpha_k^{OS}$. As expected, the overall sparsity of the PSes increases with $\alpha_k^{OS}$. As more PSes are pruned, the mean phase angle---averaged over the 155,268 PSes---and hence the static tuning power consumption decreases with increasing $\alpha_k^{OS}$. For smaller $\alpha_k^{OS}$, we observe that OS pruning provides considerable sparsity with minimal accuracy loss. In fact, with $\alpha_k^{OS}=$~2, we obtain 74.86\% PS sparsity (and a 46.05\% lower mean phase angle) with zero accuracy loss. We assume a maximum allowable accuracy loss of 5\% during pruning and therefore consider $acc_{min}^{OS}=$~91.16\%. Accordingly, the best performing OS model in our case is obtained using $\alpha_k^{OS}=$~2.4, where we achieve a sparsity of 83.77\% with a fine-tuned accuracy of 92.31\% (i.e., 3.85\% accuracy loss). Subsequently, we use this model as the starting point of the iterative pruning approach. Fig. \ref{fig3}(a) shows the simulation results for the IT approach with $\alpha_0^{IT}=$~2.4, $\Delta\alpha=$~0.2, and $acc_{min}^{IT}=$~91.16\% (5\% accuracy loss). We observe that with IT pruning, we can even achieve $\geq$~99\% sparsity with an accuracy loss less than 5\%. The best-performing IT-pruned model is obtained with $\alpha_i^{IT}=$~6 where we achieve a 99.45\% PS sparsity (and a 98.23\% reduction in mean phase angles and static power consumption) and an accuracy of 91.57\% (4.59\% accuracy loss). Fig. \ref{fig3}(b) compares the histogram of the phase angles in the unpruned, best-performing OS-pruned, and best-performing IT-pruned models. \par

We also characterize the performance of pruned models under random uncertainties in the phase angles, which is indeed critical for sparse models because even overparameterized SC-IPNNs are sensitive to such uncertainties \cite{banerjee2020modeling}. For each model, we consider 1000 Monte Carlo iterations. In each iteration, the uncertainties are sampled from a zero-mean Gaussian distribution with a standard deviation of $\sigma_{PS}\times\pi$. Fig. \ref{fig3}(c) shows the mean classification accuracy (averaged over the 1000 iterations) of the unpruned, best-performing OS-pruned, and best-performing IT-pruned models. For the pruned models, we consider two cases: 1) the pruned PSes are power-gated and left in the network (solid lines), and 2) the pruned PSes are removed from the network (dashed lines). We observe that in the first case (power-gated PSes), the network is more susceptible to phase uncertainties because even small uncertainties in otherwise zero phase angles lead to a large relative deviation in the MZI operation. In contrast, removing pruned PSes reduces the number of uncertainty-susceptible components and leads to significantly higher accuracy (up to 74\%) under uncertainties. In addition, the resulting compact network leads to a lower optical loss, and thus lower dynamic power consumption. Therefore, in situations where hardware-level modifications are feasible, the pruned PSes can be removed to improve the SC-IPNN performance under phase-shift uncertainties.\vspace{-0.07in}

\section{Conclusions}\vspace{-0.05in}
We have presented CHAMP, the first photonic hardware-aware pruning technique for SC-IPNNs. CHAMP can prune a considerable fraction of redundant PSes and increase the network sparsity by 74.86\% with no accuracy loss, 98.57\% with a 1\% accuracy loss, and 99.45\% with a 5\% accuracy loss. Executed only once per SC-IPNN, CHAMP improves the power efficiency (by up to 98.23\%) and enhances the robustness of SC-IPNNs under random uncertainties in tuned phase angles due to fabrication-process variations and thermal crosstalk.\vspace{-0.07in}
%We have presented CHAMP, the first photonic hardware-aware pruning technique for SC-IPNNs. CHAMP can prune a considerable fraction of redundant PSes and increase the network sparsity by 74.86\% with no accuracy loss, 98.57\% with a 1\% accuracy loss, and 99.45\% with a 5\% accuracy loss. This one-time optimization significantly improves the power efficiency (by up to 98.23\%) and enhances the robustness of SC-IPNNs under random uncertainties in tuned phase angles due to fabrication-process variations and thermal crosstalk.\vspace{-0.07in}

%\begin{thebibliography}{99} %% use BibTeX or add references manually
%\bibliography{sample}
%\end{thebibliography}
\bibliographystyle{unsrt}
\bibliography{sample}

\end{document}